\documentclass{article}[12pt]
\usepackage{enumerate,latexsym,fullpage}
\usepackage{amsmath,amssymb,amsthm,amsfonts,graphicx}
\usepackage{graphicx}
\usepackage{caption}
\usepackage{comment}
\usepackage{subcaption}
\usepackage[colorinlistoftodos]{todonotes}
\usepackage{flexisym}
\usepackage{hyperref}

\makeatletter
\def\munderbar#1{\underline{\sbox\tw@{$#1$}\dp\tw@\z@\box\tw@}}
\makeatother
\usepackage{mathtools}

\DeclarePairedDelimiter\abs{\lvert}{\rvert}
\makeatletter
\let\oldabs\abs
\def\abs{\@ifstar{\oldabs}{\oldabs*}}

\def\R{{\rm\vrule depth0ex width.4pt\kern-.08em R}}
\def\reals{{\rm\vrule depth0ex width.4pt\kern-.08em R}}
\def\Prob{{\rm\vrule depth0ex width.4pt\kern-.08em P}}

\newtheorem*{theorem*}{Theorem}

\newtheorem*{definition*}{Definition}

\newcommand{\mycomment}[1]{}

\makeatletter
\newcommand*\bigcdot{\mathpalette\bigcdot@{.5}}
\newcommand*\bigcdot@[2]{\mathbin{\vcenter{\hbox{\scalebox{#2}{$\m@th#1\bullet$}}}}}
\makeatother
\numberwithin{equation}{section}
\providecommand{\keywords}[1]{\textbf{\textit{Keywords:}} #1}

\newcommand\blfootnote[1]{%
  \begingroup
  \renewcommand\thefootnote{}\footnote{#1}%
  \addtocounter{footnote}{-1}%
  \endgroup
}
\usepackage{adjustbox}

\begin{document}
\title{Elasticity Based Demand Forecasting and \\ Price Optimization for Online Retail}
\author{Chengcheng Liu, M{\'a}ty{\'a}s A. Sustik\\
Walmart Labs, San Bruno, CA}
\maketitle
\begin{abstract}
We study a problem of an online retailer who observes the unit sales of a product, and dynamically changes the retail price, in order to maximize the expected revenue. Assuming the demand of the product is price sensitive, we are interested in the optimal pricing policy when future demand is uncertain. We build a system to investigate the relationship between retail price and demand and estimate the demand function. The system predicts demand and revenue at a given retail price. We formulate a revenue maximization problem over a discrete finite time horizon with discrete retail price. The optimal pricing policy is solved based on the predicted demand and revenue values. With computational experiments, we investigate the effect of optimal pricing policy to inventory management.
\end{abstract}
\blfootnote{E-mail: cl35622@utexas.edu, msustik@gmail.com}
\section{Introduction}
\label{sec: introduction}
In this paper, we assume an online retailer is selling a replenishable product, and he is able to change the retail price at any time without significant operational cost. Suppose the retailer knows the transaction history of the product. When future demand is uncertain, the retailer is interested in a pricing policy which maximizes the expected revenue for a fixed period of time. 

Price elasticity of demand is an economical measure to quantify the sensitivity of the demand of a product to changes in the retail price. Let $X$, $Y$ and $\beta$ denote the retail price, unit demand and price elasticity of demand, respectively. A general formula of price elasticity of demand is
\begin{equation*}
	\beta = \frac{\Delta Y}{\Delta X},
\end{equation*}
where $\Delta Y$ and $\Delta X$ represent the relative change of demand and retail price in percentage. Typically, we assume the price affects the demand and $\beta \neq 0$. Price elasticity is negative when the demand decreases as the price increased, and positive otherwise. 

Demand of a product is generally unknown to retailers. Bolton compares three demand models in \cite{rnb89}: a linear model, a multiplicative model and an exponential model. In this paper, we extend the multiplicative demand model by considering explanatory regression, time series effect, market value of the product, and demand autoregressive effect. We introduce the Elasticity based Demand Forecasting system to help the retailer improve his knowledge of demand uncertainty through observations of actual unit sales. For a given retail price value, the system outputs the corresponding forecasted demand and revenue values. Furthermore, we consider an optimization problem to maximize the total expected revenue within a fix period of time with an initial inventory level. We solve for an optimal pricing solution while considering the sell through rate during the time. Finally, we conduct computational experiments and compare different solutions when the sell through rate changes.

\section{Literature}
\label{sec: literature}
Revenue Management has been widely applied to retail industry especially in E-commerce retail. Overview of previous work can be found in Bittran and Caldentey (\cite{gb003}), \"{O}zer and Philips (\cite{op012}) and Talluri and Van Ryzin (\cite{tt005}). Generally, there are several aspects including marketing analysis, demand modeling, forecasting and operations research. For retailers, pricing decisions have significant effect on business profitability. A central concern is to understand the sensitivity of consumer demand to the price changes. The effect can be measured by price elasticity of demand and we highlight some previous studies in \cite{rb91}, \cite{rnb89}, \cite{cp005}, \cite{sh95} and \cite{gt88} .

There has been significant research conducted on demand modeling. Christensen and Greene apply a translog model in \cite{lc76}. Brodie and de Kluyver (\cite{rb84}) conclude that linear and multiplicative market share models perform better than the attraction model. Ben-Akiva and Lerman (\cite{ba94}) and Train (\cite{t003}) discuss the demand estimation for individual product within a set of similar products by discrete choice models, such as multinomial logit (MNL). Kalyanam (\cite{kk96}) proposes a Bayesian method for demand estimation. Vulcano and van Ryzin and Ratliff (\cite{vg012}) estimate substitute and lost demand for substitutable products based on a MNL model and a nonhomogeneous Poisson process. Talluri and van Ryzin (\cite{tk004}) develop an EM method to estimate customer arrival rate and MNL model parameters. Anupindi, Dada and Gupta (\cite{ra99}) apply the EM method to estimate demand when the first choice variant is unknown. Borle and Boatwright (\cite{bb005}) develop models of customer purchase behavior using Markov chain Monte Carlo (MCMC) techniques and analyze the impact of assortment reduction on customer retention. Kalyanam, Borle and Boatwright (\cite{kk007}) study single product demand from category sales data based on a hierarchical Bayesian framework. 

In the literature of revenue management, Bitran and Mondschein (\cite{gb97}) propose a periodic review pricing policy for selling seasonal products in retail. Caro and Gallien (\cite{cg012}) study a multi-product markdown problem for fast-fashion industry. Maddah and Bish (\cite{mb007}) analyze both pricing and inventory decisions for multiple competing products. Choi (\cite{cm007}) applies Bayesian approach to analyze inventory and pricing decisions for fashion retailers. Gallego and van Ryzin (\cite{gg94}) investigate dynamic pricing decision for a fixed time period with demand uncertainty. Petruzzi and Dada (\cite{pc002}) study optimal stocking and pricing policies with demand learning.

\section{System of Elasticity Based Demand Forecasting}
\label{sec: edf}
In this section we introduce Elasticity based Demand Forecasting (EDF) system to model the relationship between retail price and demand. The system stores multiple data sources, estimates the demand function and updates the model periodically, and predicts future demand for a given time period. Price elasticity is estimated along with the demand model. Demand forecasting values are further utilized in the optimal pricing solution in Section \ref{sec: opt}.
\subsection{System Structure}
\begin{figure}[h!]
\centering

\tikzset{every picture/.style={line width=0.75pt}} 

\begin{tikzpicture}[x=0.75pt,y=0.75pt,yscale=-1,xscale=1]

\draw   (36.33,260.66) .. controls (36.33,254.62) and (41.23,249.73) .. (47.27,249.73) -- (131.4,249.73) .. controls (137.44,249.73) and (142.33,254.62) .. (142.33,260.66) -- (142.33,293.46) .. controls (142.33,299.5) and (137.44,304.4) .. (131.4,304.4) -- (47.27,304.4) .. controls (41.23,304.4) and (36.33,299.5) .. (36.33,293.46) -- cycle ;
\draw    (143.33,270.4) -- (174,270.49) ;
\draw [shift={(176,270.5)}, rotate = 180.18] [color={rgb, 255:red, 0; green, 0; blue, 0 }  ][line width=0.75]    (10.93,-3.29) .. controls (6.95,-1.4) and (3.31,-0.3) .. (0,0) .. controls (3.31,0.3) and (6.95,1.4) .. (10.93,3.29)   ;
\draw    (6.15,363.2) -- (6.33,512.23) ;
\draw   (24.96,350.36) .. controls (24.96,346.15) and (28.38,342.73) .. (32.59,342.73) -- (161,342.73) -- (161,380.86) -- (24.96,380.86) -- cycle ;
\draw   (24.96,402.53) .. controls (24.96,398.32) and (28.38,394.9) .. (32.59,394.9) -- (163,394.9) -- (163,433.03) -- (24.96,433.03) -- cycle ;
\draw   (24.96,453.1) .. controls (24.96,448.88) and (28.38,445.47) .. (32.59,445.47) -- (164,445.47) -- (164,483.6) -- (24.96,483.6) -- cycle ;
\draw   (25.75,502.86) .. controls (25.75,498.65) and (29.16,495.24) .. (33.37,495.24) -- (161,495.24) -- (161,533.36) -- (25.75,533.36) -- cycle ;
\draw    (180.15,360.9) -- (180.33,509.93) ;
\draw    (161,360.89) -- (180.15,360.9) ;
\draw    (161.18,509.93) -- (180.33,509.93) ;
\draw    (367,250.5) -- (366.94,220.32) ;
\draw [shift={(366.94,218.32)}, rotate = 449.89] [color={rgb, 255:red, 0; green, 0; blue, 0 }  ][line width=0.75]    (10.93,-3.29) .. controls (6.95,-1.4) and (3.31,-0.3) .. (0,0) .. controls (3.31,0.3) and (6.95,1.4) .. (10.93,3.29)   ;
\draw    (550.67,306.23) -- (550.98,337.5) ;
\draw [shift={(551,339.5)}, rotate = 269.43] [color={rgb, 255:red, 0; green, 0; blue, 0 }  ][line width=0.75]    (10.93,-3.29) .. controls (6.95,-1.4) and (3.31,-0.3) .. (0,0) .. controls (3.31,0.3) and (6.95,1.4) .. (10.93,3.29)   ;
\draw    (89.67,221.23) -- (89.92,246.32) ;
\draw [shift={(89.94,248.32)}, rotate = 269.43] [color={rgb, 255:red, 0; green, 0; blue, 0 }  ][line width=0.75]    (10.93,-3.29) .. controls (6.95,-1.4) and (3.31,-0.3) .. (0,0) .. controls (3.31,0.3) and (6.95,1.4) .. (10.93,3.29)   ;
\draw    (89,156) -- (89.25,181.09) ;
\draw [shift={(89.27,183.09)}, rotate = 269.43] [color={rgb, 255:red, 0; green, 0; blue, 0 }  ][line width=0.75]    (10.93,-3.29) .. controls (6.95,-1.4) and (3.31,-0.3) .. (0,0) .. controls (3.31,0.3) and (6.95,1.4) .. (10.93,3.29)   ;
\draw   (399,30) -- (399,82) .. controls (399,86.97) and (385.57,91) .. (369,91) .. controls (352.43,91) and (339,86.97) .. (339,82) -- (339,30) .. controls (339,25.03) and (352.43,21) .. (369,21) .. controls (385.57,21) and (399,25.03) .. (399,30) .. controls (399,34.97) and (385.57,39) .. (369,39) .. controls (352.43,39) and (339,34.97) .. (339,30) ;
\draw   (89.5,100) -- (174,128.5) -- (89.5,157) -- (5,128.5) -- cycle ;
\draw    (6.15,363.2) -- (25.3,363.2) ;
\draw    (6.33,512.23) -- (25.49,512.23) ;
\draw   (76,319.7) -- (90.5,304.5) -- (105,319.7) -- (97.75,319.7) -- (97.75,342.5) -- (83.25,342.5) -- (83.25,319.7) -- cycle ;
\draw   (295.85,190.66) .. controls (295.85,186.76) and (299.01,183.6) .. (302.91,183.6) -- (429.94,183.6) .. controls (433.84,183.6) and (437,186.76) .. (437,190.66) -- (437,211.84) .. controls (437,215.74) and (433.84,218.9) .. (429.94,218.9) -- (302.91,218.9) .. controls (299.01,218.9) and (295.85,215.74) .. (295.85,211.84) -- cycle ;
\draw    (640.17,382.22) -- (681,382.5) ;
\draw    (89,57.5) -- (339,57.5) ;
\draw    (89,57.5) -- (88.52,98) ;
\draw [shift={(88.5,100)}, rotate = 270.67] [color={rgb, 255:red, 0; green, 0; blue, 0 }  ][line width=0.75]    (10.93,-3.29) .. controls (6.95,-1.4) and (3.31,-0.3) .. (0,0) .. controls (3.31,0.3) and (6.95,1.4) .. (10.93,3.29)   ;
\draw    (425.33,268.4) -- (496,268.5) ;
\draw [shift={(498,268.5)}, rotate = 180.08] [color={rgb, 255:red, 0; green, 0; blue, 0 }  ][line width=0.75]    (10.93,-3.29) .. controls (6.95,-1.4) and (3.31,-0.3) .. (0,0) .. controls (3.31,0.3) and (6.95,1.4) .. (10.93,3.29)   ;
\draw    (463.15,361.2) -- (463,410.5) ;
\draw   (481.96,348.36) .. controls (481.96,344.15) and (485.38,340.73) .. (489.59,340.73) -- (618,340.73) -- (618,378.86) -- (481.96,378.86) -- cycle ;
\draw   (481.96,400.53) .. controls (481.96,396.32) and (485.38,392.9) .. (489.59,392.9) -- (620,392.9) -- (620,431.03) -- (481.96,431.03) -- cycle ;
\draw    (640,358.5) -- (640.33,407.93) ;
\draw    (618,358.89) -- (640,358.5) ;
\draw    (621.18,407.93) -- (640.33,407.93) ;
\draw    (463.15,361.2) -- (482.3,361.2) ;
\draw    (463,410.5) -- (482.15,410.5) ;
\draw   (177.33,261.66) .. controls (177.33,255.62) and (182.23,250.73) .. (188.27,250.73) -- (272.4,250.73) .. controls (278.44,250.73) and (283.33,255.62) .. (283.33,261.66) -- (283.33,294.46) .. controls (283.33,300.5) and (278.44,305.4) .. (272.4,305.4) -- (188.27,305.4) .. controls (182.23,305.4) and (177.33,300.5) .. (177.33,294.46) -- cycle ;
\draw    (284.33,271.4) -- (315,271.49) ;
\draw [shift={(317,271.5)}, rotate = 180.18] [color={rgb, 255:red, 0; green, 0; blue, 0 }  ][line width=0.75]    (10.93,-3.29) .. controls (6.95,-1.4) and (3.31,-0.3) .. (0,0) .. controls (3.31,0.3) and (6.95,1.4) .. (10.93,3.29)   ;
\draw    (369,183.5) -- (369,92.5) ;
\draw [shift={(369,90.5)}, rotate = 450] [color={rgb, 255:red, 0; green, 0; blue, 0 }  ][line width=0.75]    (10.93,-3.29) .. controls (6.95,-1.4) and (3.31,-0.3) .. (0,0) .. controls (3.31,0.3) and (6.95,1.4) .. (10.93,3.29)   ;
\draw    (681,382.5) -- (681,56.5) ;
\draw    (681,56.5) -- (403,57.49) ;
\draw [shift={(401,57.5)}, rotate = 359.8] [color={rgb, 255:red, 0; green, 0; blue, 0 }  ][line width=0.75]    (10.93,-3.29) .. controls (6.95,-1.4) and (3.31,-0.3) .. (0,0) .. controls (3.31,0.3) and (6.95,1.4) .. (10.93,3.29)   ;
\draw   (18.85,192.66) .. controls (18.85,188.76) and (22.01,185.6) .. (25.91,185.6) -- (152.94,185.6) .. controls (156.84,185.6) and (160,188.76) .. (160,192.66) -- (160,213.84) .. controls (160,217.74) and (156.84,220.9) .. (152.94,220.9) -- (25.91,220.9) .. controls (22.01,220.9) and (18.85,217.74) .. (18.85,213.84) -- cycle ;
\draw   (317.33,261.66) .. controls (317.33,255.62) and (322.23,250.73) .. (328.27,250.73) -- (412.4,250.73) .. controls (418.44,250.73) and (423.33,255.62) .. (423.33,261.66) -- (423.33,294.46) .. controls (423.33,300.5) and (418.44,305.4) .. (412.4,305.4) -- (328.27,305.4) .. controls (322.23,305.4) and (317.33,300.5) .. (317.33,294.46) -- cycle ;
\draw   (497.33,262.66) .. controls (497.33,256.62) and (502.23,251.73) .. (508.27,251.73) -- (592.4,251.73) .. controls (598.44,251.73) and (603.33,256.62) .. (603.33,262.66) -- (603.33,295.46) .. controls (603.33,301.5) and (598.44,306.4) .. (592.4,306.4) -- (508.27,306.4) .. controls (502.23,306.4) and (497.33,301.5) .. (497.33,295.46) -- cycle ;

\draw (89.5,128.5) node   [align=left] {\begin{minipage}[lt]{102.15pt}\setlength\topsep{0pt}
\begin{center}
Product Eligibility
\end{center}

\end{minipage}};
\draw (89.33,276.06) node   [align=left] {\begin{minipage}[lt]{86.72pt}\setlength\topsep{0pt}
\begin{center}
Data\\Pre-processing
\end{center}

\end{minipage}};
\draw (369.75,65.74) node   [align=left] {\begin{minipage}[lt]{66.98pt}\setlength\topsep{0pt}
\begin{center}
Data \\Storage\\
\end{center}

\end{minipage}};
\draw (92.71,362.5) node   [align=left] {\begin{minipage}[lt]{90.16pt}\setlength\topsep{0pt}
\begin{center}
Data Aggregation
\end{center}

\end{minipage}};
\draw (93.98,415.09) node   [align=left] {\begin{minipage}[lt]{85.05pt}\setlength\topsep{0pt}
\begin{center}
Missing Data
\end{center}

\end{minipage}};
\draw (94.48,464.53) node   [align=left] {\begin{minipage}[lt]{89.48pt}\setlength\topsep{0pt}
\begin{center}
Transformation \&\\Normalization
\end{center}

\end{minipage}};
\draw (91.31,515.28) node   [align=left] {\begin{minipage}[lt]{93.41pt}\setlength\topsep{0pt}
\begin{center}
Outlier Detection
\end{center}

\end{minipage}};
\draw (366.42,201.56) node   [align=left] {\begin{minipage}[lt]{101.32pt}\setlength\topsep{0pt}
\begin{center}
Evaluation
\end{center}

\end{minipage}};
\draw (550.15,360.83) node   [align=left] {\begin{minipage}[lt]{92.27pt}\setlength\topsep{0pt}
\begin{center}
Elasticity 
\end{center}

\end{minipage}};
\draw (553.72,411.7) node   [align=left] {\begin{minipage}[lt]{90.14pt}\setlength\topsep{0pt}
\begin{center}
Demand Forecast
\end{center}

\end{minipage}};
\draw (230.33,277.06) node   [align=left] {\begin{minipage}[lt]{86.72pt}\setlength\topsep{0pt}
\begin{center}
Feature\\Extraction
\end{center}

\end{minipage}};
\draw (89.42,202.25) node   [align=left] {\begin{minipage}[lt]{101.32pt}\setlength\topsep{0pt}
\begin{center}
Data Input
\end{center}

\end{minipage}};
\draw (370.33,278.06) node   [align=left] {\begin{minipage}[lt]{86.72pt}\setlength\topsep{0pt}
\begin{center}
Model
\end{center}

\end{minipage}};
\draw (550.33,279.06) node   [align=left] {\begin{minipage}[lt]{86.72pt}\setlength\topsep{0pt}
\begin{center}
Prediction
\end{center}

\end{minipage}};

\end{tikzpicture}
\caption{EDF System} \label{fig:system structure}
\end{figure}
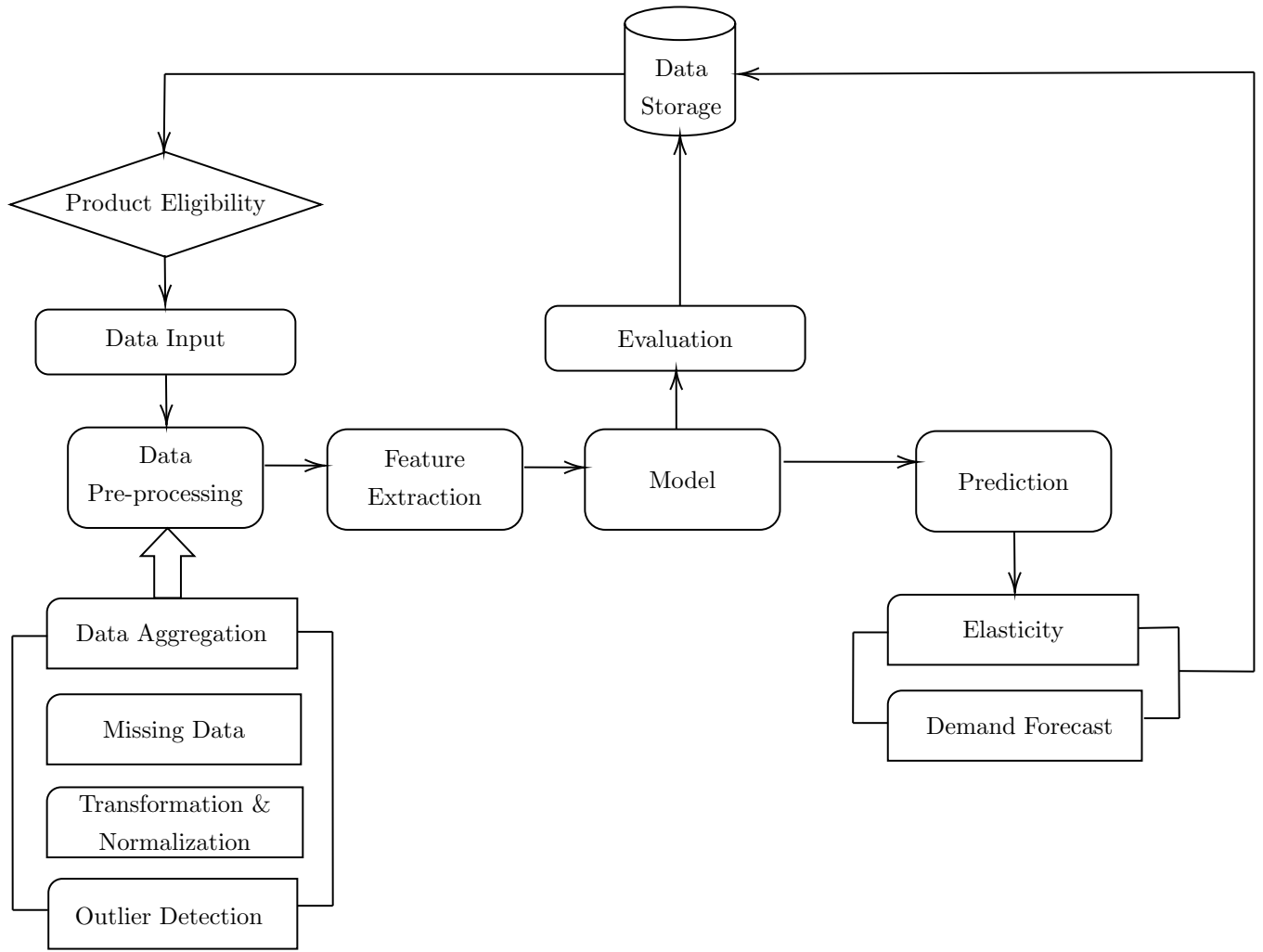
The system structure is shown in figure \ref{fig:system structure}. There are four major components in the system: data storage, data preparation, model processing and prediction. 

Data storage contains multiple types of data: product description data, product market value data, model related data and prediction data. Product description data describe the product information such as brand and category etc. Historical transaction data describe online transaction details with timestamps, for example, retail price, discount amount and purchase quantity etc. Product market value data describe the market value for the product. Assuming the retailer observes retail prices offered by some other online retailers who sell the same product, we keep the minimum value of the competitor prices for further feature preparation. Demand model is estimated based on the aforementioned data. Model related data include price elasticities, model information and evaluation metrics. When the model is used for prediction, prediction data include the forecasted demand values for a given retail price over a given time horizon. The system reads and writes data on a regular basis to keep the data and models updated.

Data preparation component has multiple modules: Product Eligibility, Data Pre-processing and Feature Extraction. Product Eligibility module is designed to select a set of products which are eligible to enter the system. The selection is processed based on several configurable rules, and the rules are defined by the system manager. For each product, the system examines both historical transaction data and model related data. For a given time period, the module counts the number of days with transactions and number of distinct retail prices. If model related data are available, the module looks for the latest model, and finds the corresponding estimated price elasticity, confidence level of the elasticity estimation, and demand model evaluation metrics. The module further compares the results with the predefined threshold values from the system configuration. A product is considered to be eligible only if the results are no worse than the thresholds. The module applies these rules in order to better control the system output quality. Data Pre-processing module integrates data aggregation, missing data processing, data transformation, data normalization and outlier detection. Generally, historical transaction data has timestamps in date-time format. Data aggregation step aggregates the transaction data into daily data. For example, daily sales are the sum of hourly sales, and daily price is the average price weighted by hourly sales. Missing data processing, data transformation and data normalization are general steps in order to preprocess the original data. Outlier detection step checks the extreme values of sales. Feature Extraction module prepares the model features after the Data pre-processing module is complete.

Model processing component receives the prepared features as input, and trains the demand model for a single product. The model quality is evaluated and the evaluation metrics are stored in the data storage. The retailer manages the operation schedule of the system, to refresh the data storage and update the demand models for every eligible product. For each product, there might be multiple demand models, and each model has a unique version number. Prediction component refers to the latest models when there are multiple versions. Price elasticity is an estimated coefficient of the demand model. The demand model generates predictions of future demand for a given time period. The output results are saved in data storage, and are used in Product Eligibility module and the price recommendation problem Section \ref{sec: opt}.

\subsection{Data and Input Preparation}
Demand model estimation in the system requires both historical transaction data and product market value data to prepare the model input. Historical transaction data are time series data with a timestamp filed for a period of time. We first aggregate the original hourly transactions into daily unit sales and retail prices. If there are different retail prices during a single day from multiple online transactions, the daily retail price is the average retail price weighted by the corresponding unit sales. Secondly, base on the date information in the data, we extract two additional binary features: $is\_holiday$ and $is\_weekend$. If a single day is a national holiday or a weekend day, the corresponding feature has a value of 1, and 0 otherwise. 

Product market value data collect retail prices offered by other online retailers for the same product. The data are first transformed into daily minimum values $min\_other\_price$ to represent the most competitive market price of the product. Secondly, we define a feature $competitive\_indicator$ and
\begin{equation*}
	competitive\_indicator = \frac{retail\_price}{retail\_price + min\_other\_price},
\end{equation*}
where $retail\_price$ is the price offered by the retailer. The feature $competitive\_indicator$ is a value between 0 and 1, and a smaller value indicates the retail price is more competitive in the market.


\subsection{Model}
We follow the price elasticity definition in Section \ref{sec: introduction} and consider multiple factors that affect the demand. We assume the demand is a function of long-term trend, seasonality, recent sales and explanatory features including $retail\_price$, $competitive\_indicator$, $is\_holiday$ and $is\_weekend$. Let $X$ denote the explanatory features, $L$ and $S$ denote the long-term trend and seasonality, and $U$ denote the recent sales. Equivalently,
\begin{equation*}
	Y = a^{T}X+L+S+U,
\end{equation*}
where $a$ represents the vector of coefficients, and $X$ represents the vector of explanatory features.

Specifically, we introduce the following notation:\\
\begin{tabular}{ll}
$y_{t}$ & demand of time $t$ \\
$\tilde{y}$ & average demand \\
$x_{t}$ & retail price of time $t$ \\
$\tilde{x}$ & average retail price \\
$\beta_{x}$ & price elasticity of demand \\
$c_{t}$ & competitive indicator of time $t$ \\
$\beta_{c}$ & coefficient of competitive indicator \\
$h_{t}$ & 1 if time $t$ is a holiday, 0 otherwise \\
$\beta_{h}$ & coefficient of holiday effect \\
$w_{t}$ & 1 if time $t$ is a weekend, 0 otherwise \\
$\beta_{w}$ & coefficient of weekend effect \\
$\mu_{t}$ & trend of time $t$\\
$s_{t}$ & seasonality of time $t$ \\
$\rho$ & autoregressive coefficient
\end{tabular}\\
\begin{align}
	\log \frac{y_{t}}{\tilde{y}} &= \beta_{x} \log \frac{x_{t}}{\tilde{x}} + \beta_{c}c_{t}+ \beta_{h} h_{t} + \beta_{w} w_{t} + \mu_{t} + s_{t} + \epsilon_{t} \label{model} \\
	\mu_{t} &= \mu_{t-1} + \gamma_{t-1} \label{trend1} \\
	\gamma_{t} &= \gamma_{t-1} + \tau_{t} \label{trend2} \\
	\tau_{t} &\sim N(0, \sigma^{2}_{\tau}) \label{trend3} \\
	s_{t} &= -\sum_{j=1}^{k-1}s_{t+1-j} + \omega_{t} \label{sea1} \\
	\omega_{t} &\sim N(0, \sigma^{2}_{\omega}) \label{sea2} \\
	\epsilon_{t} &= \rho \epsilon_{t-1} + \eta_{t} \label{ar1} \\
	\eta_{t} &\sim N(0, \sigma^{2}_{\eta}) \label{ar2}
\end{align}
In model \ref{model}, $x_{t}$, $c_{t}$, $h_{t}$ and $w_{t}$ represent the retail price, competitive indicator, holiday binary feature and weekend binary feature, respectively, and price elasticity of demand is represented by $\beta_{x}$. We use $\tilde{y}$ and $\tilde{x}$ to normalize the retail price and unit sales. And they are calculated based on historical data. \ref{trend1}, \ref{trend2} and \ref{trend3} model the long-term smooth trend. \ref{sea1} and \ref{sea2} model the seasonality, and $k$ is the periodicity. \ref{ar1} and \ref{ar2} model the autoregressive effect from previous sales to future demand. We assume the autoregressive degree is 1 and the coefficient is denoted by $\rho$. The model is estimated based on daily time series data. 

\section{Revenue Optimization and Price Recommendation}
\label{sec: opt}
\subsection{Problem Setup}
\label{sec: opt setup}
In this section, we are interested in the optimal pricing policy for the retailer to maximize the total expected revenue of a single product over a finite time period. The time period may have a length of a number of days, weeks or months. We assume there is a given inventory level at the beginning of the time period. 

First, instead of modeling the retail price as a continuous variable in the optimization problem, we model it as a discrete variable. The variable support is defined by the historical minimum and maximum retail price values, and the discrete values are evenly distributed within the range. We introduce a set of price levels to represent each discrete price value.

Second, we formulate the problem in a discrete time horizon, and the retail price may change at each time point. For example, if the retailer is planning for a 10 day length time period, and we assume each day there is a single retail price, and we do not allow multiple prices during a single day in this case.

Third, the retailer is interested in planning for a pricing policy for future, and the demand is uncertain during the given time period. According to Section \ref{sec: edf}, we assume the retailer has the estimated demand function from the EDF system. For each discrete price value in the predefined set of retail prices, we are able to get the corresponding predicted demand value. Furthermore, we multiply the discrete price values and the predicted demand values to get the corresponding predicted revenue values. 

Furthermore, we define a sell through rate during the time period as the number of total sales divided by the starting inventory. For example, if there are 100 units in the beginning of a month, and there are 20 units sold within the month, then the sell through rate during the month is $0.2$. On the other hand, If we have a sell through rate great than or equal to $0.2$, then there are at least 20 units sold, or equivalently, at most 80 units left at the end of the month.

We formulate an optimization problem of which the objective is the total expected revenue over the time period. We put a minimum threshold for the sell through rate in the constraint set, in order to control the remaining inventory when the time period ends. 
\subsection{Problem Formulation}
We first introduce the mathematical notation of the optimization problem.\\
Sets\\
\begin{tabular}{ll}
$I = |1, 2, \cdots, k |$ & set of price levels\\
$T = |1, 2, \cdots, n |$ & set of time points, such as week 1, week 2, etc.
\end{tabular}\\
Parameters\\
\begin{tabular}{lll}
$p_{i}$ & $i \in I$ & retail price at level $i$ \\
$d_{it}$ & $i \in I, t \in T$ & demand prediction at price level $i$ at time $t$ \\
$r_{it}$ & $i \in I, t \in T$ & revenue prediction at price level $i$ at time $t$ \\
$s_{0}$ && starting inventory level \\
$\alpha$ && minimum sell through rate \\
\end{tabular}\\
Decision Variables\\
\begin{tabular}{lll}
$x_{it}$ & $i \in I, t \in T$ & 1 if price level $i$ is selected at time $t$, 0 otherwise\\
$s_{t}$ & $t \in T$ & remaining inventory at the end of time $t$ \\
\end{tabular}\\
\setcounter{equation}{0}
\begin{align}
\max \quad
& \sum_{t \in T}\sum_{i \in I}r_{it}x_{it} \label{obj}
\end{align}
Subject to
\begin{align}
\sum_{i \in I} x_{it} &= 1 && \forall t \in T \label{op1} \\
s_{t-1} &= \sum_{i \in I} d_{it} x_{it} + s_{t} && \forall t \in T \label{op2} \\
s_{n} &\leq (1-\alpha) s_{0} \label{op3} \\
s_{t} &\geq 0 && \forall t \in T\\
x_{it} &\in \{0, 1\}&& \forall i\in I, t \in T
\end{align}
We define a price level variable $x_{it} = 1$ if price level $i$ is selected at time $t$, and $x_{it} = 0$ otherwise. $s_{t}$ is the inventory variable at the end of time $t$. The total expected revenue over the given time horizon is represented by \ref{obj}. Constraint \ref{op1} says only one price level is chosen at each time point. \ref{op2} models the demand and remaining inventory at each time point, assuming there is no replenishment during the whole time period. If $\alpha$ represents the minimum sell through rate, constraint \ref{op3} limits the remaining inventory by not exceeding $(1-\alpha)s_{0}$, where $s_{0}$ is the inventory level in the beginning. We are interested in an optimal pricing policy with at least a certain level of inventory sold during the time period.
\newpage
\section{Computational Results}
\label{sec: computational}
In this section we study an example of a single product. The input data cover the recent two-year history. We define the optimization time period as the future eight consecutive weeks, so each time point represents a week. Assuming that the retailer only changes the retail price per week, the optimal pricing solution should have eight weekly price values in this example. 

We first establish the demand model estimation through the EDF system with daily time series data. Then we find the historical minimum and maximum retail prices and define the set of price values. For each discrete price value in the set, we get the forecasted demand and revenue values on a daily basis. The daily forecasted values are further aggregated into weekly forecasted values. The optimization problem receives the input of weekly forecasted demand and revenues.

\begin{figure}[h!]
\centering
\includegraphics[width=\textwidth,height=\textheight,keepaspectratio]{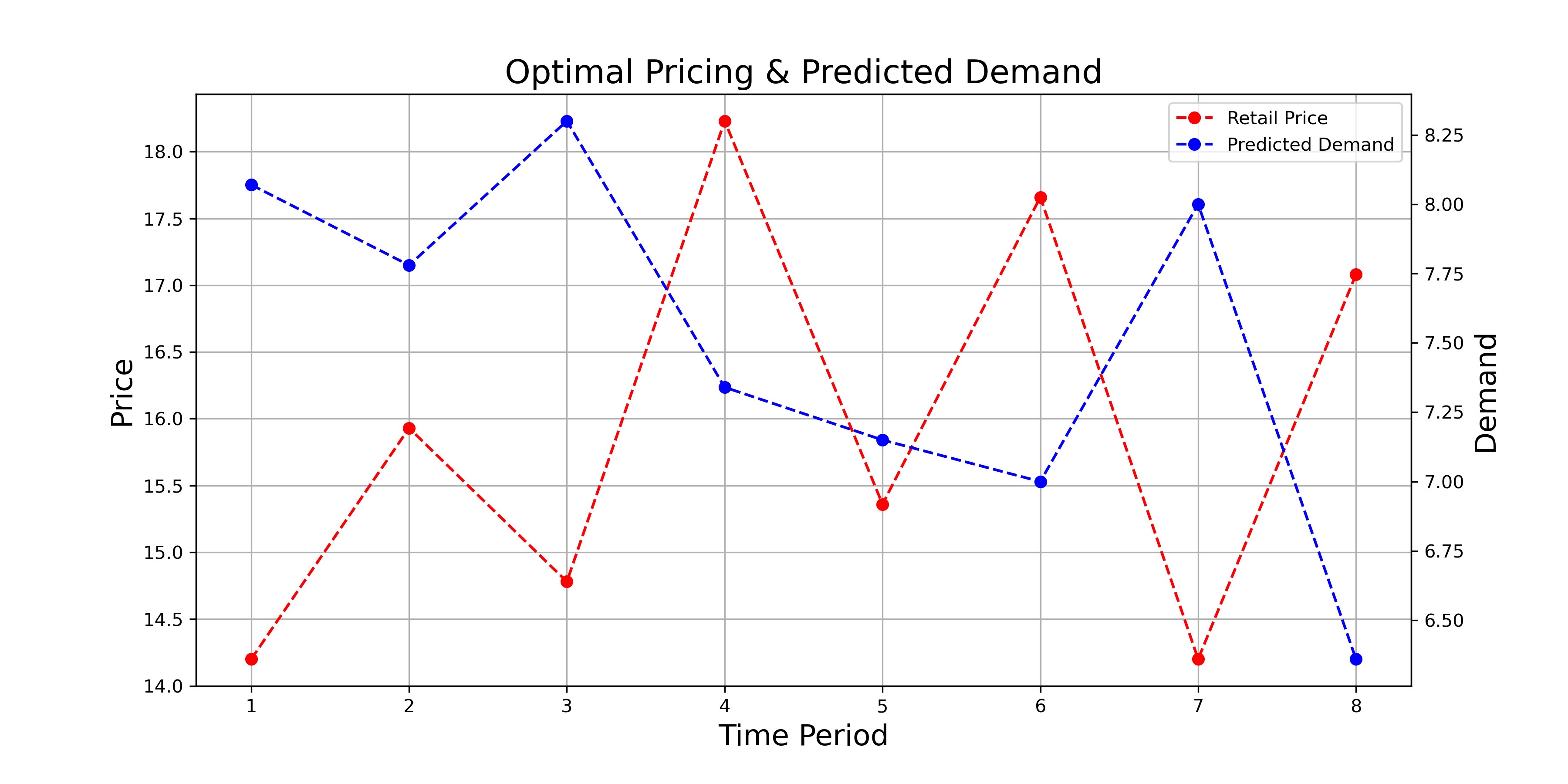}
\captionsetup{justification=raggedright,singlelinecheck=on}
\caption{Example: Optimal Pricing and Predicted Demand}\label{fig: pred demand}
\end{figure}

\begin{figure}[h!]
\centering
\includegraphics[width=\textwidth,height=\textheight,keepaspectratio]{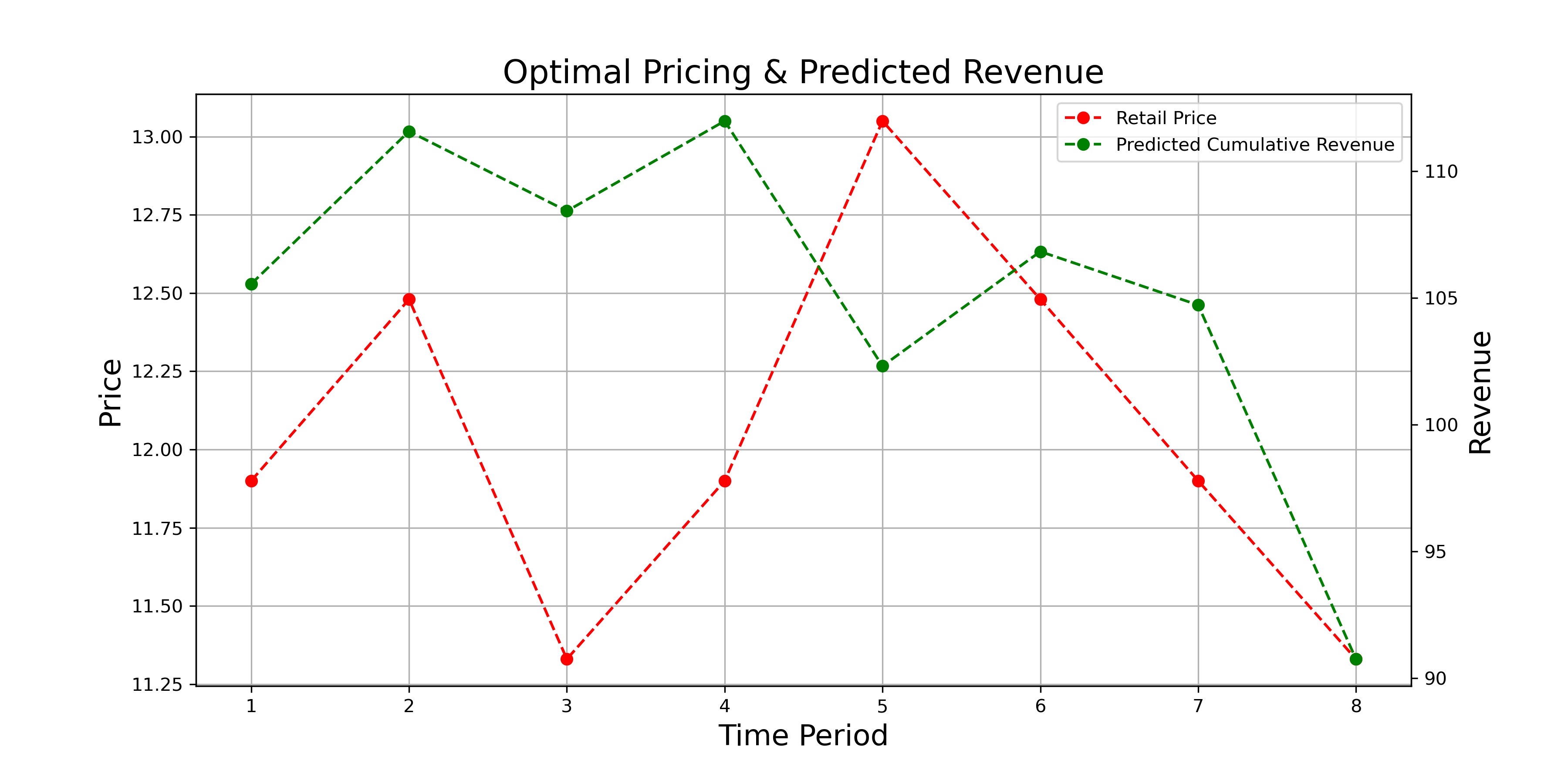}
\captionsetup{justification=raggedright,singlelinecheck=on}
\caption{Example: Optimal Pricing and Predicted Revenue}\label{fig: pred revenue}
\end{figure}

We present some results with different settings for the minimum sell through rate $\alpha$. In figure \ref{fig: pred demand} and \ref{fig: pred revenue}, we solve for the optimal pricing policy when $\alpha=0.4$. Figure \ref{fig: pred demand} shows the optimal pricing policy and the predicted demand for each week. Figure \ref{fig: pred revenue} shows the corresponding predicted revenues for each week.
\newpage
Next we study the effect of sell through rate to the optimal pricing solution. With the definition of sell through rate in Section \ref{sec: opt setup}, as $\alpha$ increases, the remaining inventory at the end of the time period decreases. We let $\alpha$ varies between 0.4 and 0.7, and solve for the optimal pricing solutions accordingly. 

Figure \ref{fig: pred revenue} shows the change of optimal pricing solution as $\alpha$ increases. We observe that as $\alpha$ increases, the optimal retail price decreases. Intuitively, if the price elasticity of demand is negative, as the retail price decreases, the demand increases, thus the sell through rate increases.
\begin{figure}[h!]
\centering
\includegraphics[width=\textwidth,height=\textheight,keepaspectratio]{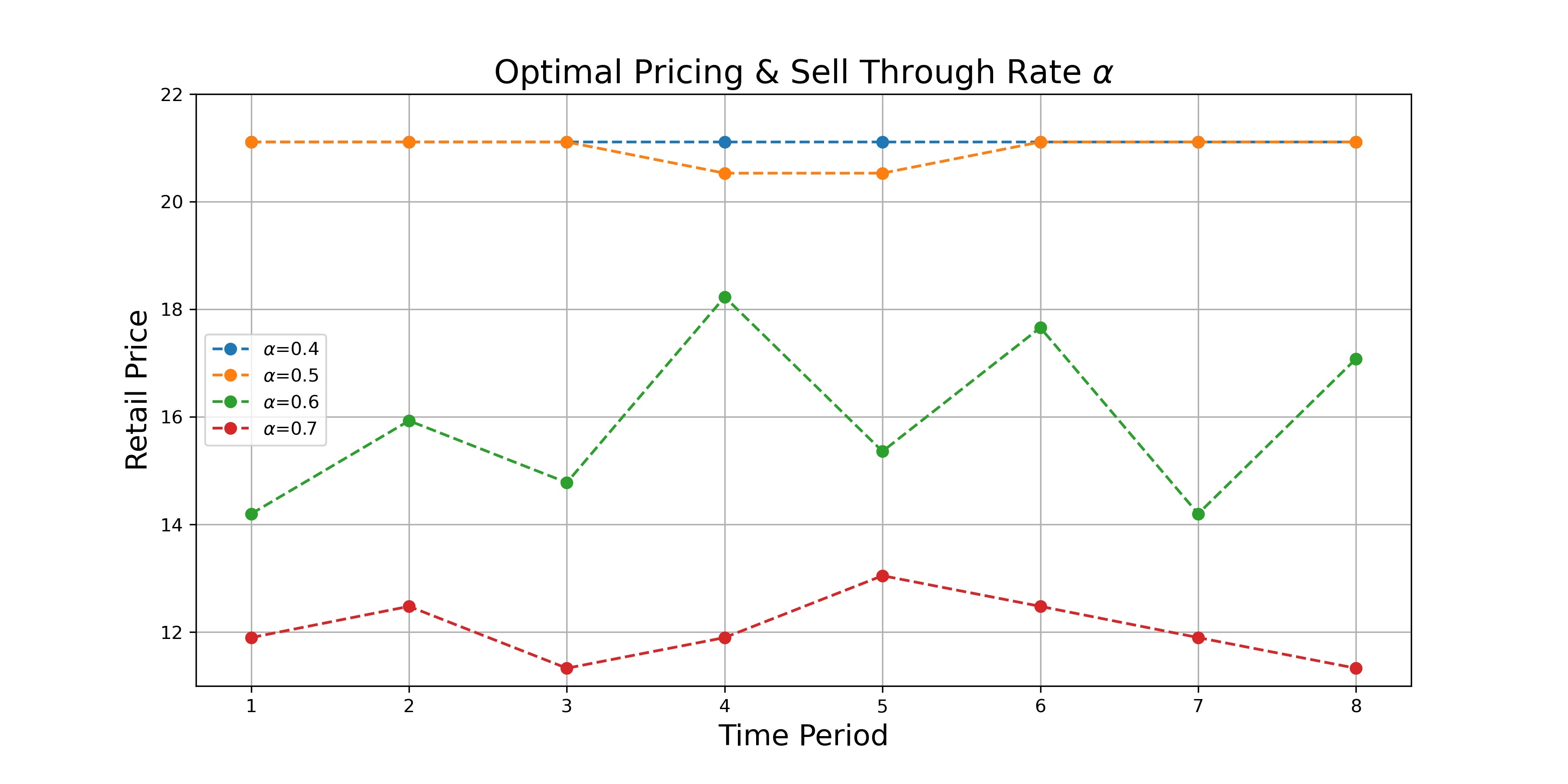}
\captionsetup{justification=raggedright,singlelinecheck=on}
\caption{Example: Optimal Pricing and Sell Through Rate}\label{fig: op str}
\end{figure}

\begin{figure}[h!]
\centering
\includegraphics[width=\textwidth,height=\textheight,keepaspectratio]{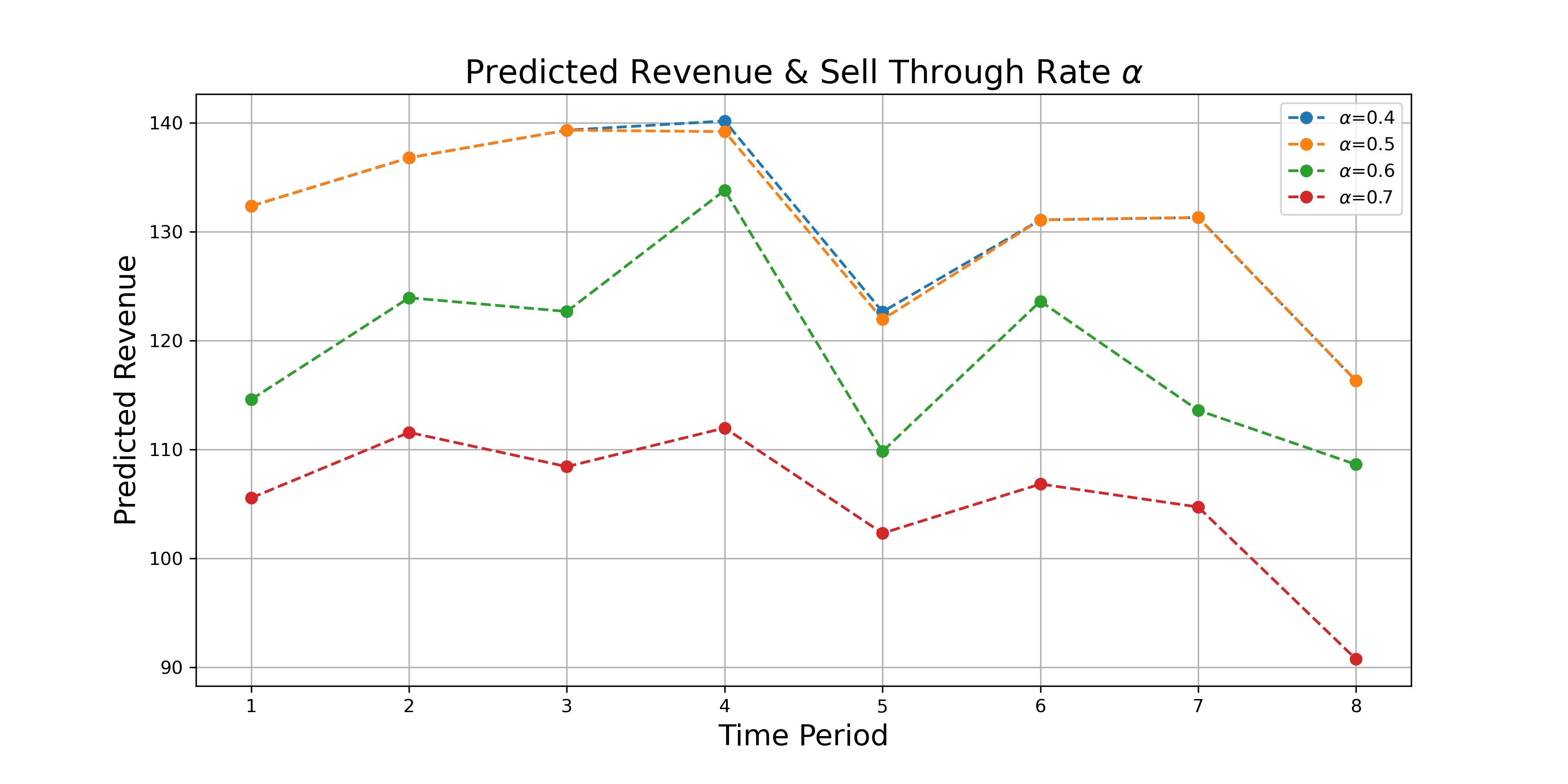}
\captionsetup{justification=raggedright,singlelinecheck=on}
\caption{Example: Predicted Revenue and Sell Through Rate}\label{fig: pred revenue str}
\end{figure}

Furthermore, figure \ref{fig: pred revenue str} shows the predicted revenues as $\alpha$ changes. In this case, the revenue decreases when $\alpha$ increases. We observed that for this example product, when there is a large chance of overstocking, the retailer should decrease the retail price. For example, if the product is a seasonal product, and is only available to sell during holiday seasons, the retailer would prefer less remaining inventory after the holiday seasons end. When the business goal is to increase the revenue over a time period without significant inventory concern, the retailer should increase the retail price.
\newpage
\section{Conclusions}
In this paper, we study the optimal pricing decision for an online retailer when the product demand is uncertain. Based on the definition of price elasticity, we assume that demand is driven by retail price, price competitive indicator, holidays, weekends, long-term trend, seasonality and demand autoregressive effect. We introduce the Elasticity based Demand Forecasting system and propose a systematic method to select eligible products, process input data, estimate demand model and predict future demand. The retailer manages the system configuration and operation schedule. The system automatically identifies products that satisfy the model assumptions, estimates the demand models and outputs the forecasting results. 

We then formulate a discrete-time revenue-maximization problem over a finite time period. We define a sell through rate to measure the inventory management. The retail price decision variable is modeled as a discrete variable with finite values. We conduct computational experiments to study the effect of the optimal pricing policy to the sell through inventory. 
\newpage


\bibliographystyle{abbrv}  
\bibliography{edf}        
\index{Bibliography@\emph{Bibliography}}
\end{document}